\documentclass[aps,prb,amsfonts,amssymb,amsmath,groupedaddress,twocolumn]{revtex4-2}

\usepackage{graphicx}
\usepackage{amsmath}
\usepackage{amsfonts}
\usepackage{amssymb}
\usepackage{graphicx}
\usepackage{epstopdf}

\begin{document}

\title{The Limits of Thermoelectric Performance with a Bounded Transport Distribution}

\author{Jesse Maassen}
\email{jmaassen@dal.ca}
\affiliation{Department of Physics and Atmospheric Science, Dalhousie University, Halifax, Nova Scotia, Canada, B3H 4R2}

\begin{abstract}
With the goal of maximizing the thermoelectric (TE) figure of merit $ZT$, Mahan and Sofo [Proc. Natl. Acad. Sci. U.S.A. {\bf 93}, 7436 (1996)] found that the optimal transport distribution (TD) is a delta function. Materials, however, have TDs that appear to always be finite and non-diverging. Motivated by this observation, this study focuses on deriving what is the optimal {\it bounded} TD, which is determined to be a boxcar function for $ZT$ and a Heaviside function for power factor. From these optimal TDs upper limits on $ZT$ and power factor are obtained; the maximum $ZT$ scales with $\Sigma_{\rm max} T /\kappa_l$, where $\Sigma_{\rm max}$ is the TD magnitude and $\kappa_l$ is the lattice thermal conductivity. These results help establish practical upper limits on the performance of TE materials and provide target TDs to guide band/scattering engineering strategies.
\end{abstract}

\maketitle

\section{Introduction}
Thermoelectric (TE) materials enable the interconversion of electrical and thermal energy, and are being explored for applications in power generation and solid-state cooling. The TE efficiency is closely related to the TE figure of merit $ZT$\,=\,$S^2\sigma T$/($\kappa_e$+$\kappa_l$), where $S$ is the Seebeck coefficient, $\sigma$ is the electrical conductivity, $\kappa_{e,l}$ are the electronic/lattice thermal conductivities, and $T$ is the temperature \cite{Snyder2008}. As $ZT$ increases so does the TE efficiency, which, in principle, can reach the upper limit Carnot efficiency as $ZT$ approaches infinity. The TE properties related to electron transport, namely $\sigma$, $S$ and $\kappa_e$, are determined by a central quantity known as the transport distribution (TD), which is unique to each material \cite{Mahan1996}.

Many strategies to improve the electronic component of TEs have been proposed and demonstrated \cite{Pei2012}, such as distortion of the electronic states \cite{Heremans2008}, band convergence \cite{Pei2011}, TEs with low $\kappa_e$ \cite{McKinney2017}, low dimensional materials \cite{Hicks1993,Kim2009}, TEs with reduced/optimized scattering \cite{Shuai2017,Mao2017,Zhou2018}, semimetals \cite{Markov2018,Markov2019}, narrow-gap semiconductors \cite{Graziosi2020} and unusual band shapes \cite{Zahid2010,Chen2013,Maassen2013,Wickramaratne2015,Rudderham2021}. These approaches, ultimately, have the effect of altering the TD with the goal of enhancing $ZT$ and/or the power factor $PF$\,=\,$S^2\sigma$. It is then important to know what is the best TD for TE performance. This was the focus of the seminal work by Mahan and Sofo \cite{Mahan1996}, which concluded that the optimal $ZT$ originates from a delta function TD of the form $\Sigma_{\rm MS}\, \delta(x$$-$$x'')$, where $\Sigma_{\rm MS}$ is a constant, $x$\,=\,($E$$-$$\mu$)/$k_B T$ is the normalized electron energy, $\mu$ is the chemical potential and $x''$\,=\,2.40. This TD, by design, has an infinitely narrow width that makes $\kappa_e$\,=\,0 to enhance $ZT$, but gives a finite $\sigma$ and $PF$ since it diverges to infinity when $x$\,=\,$x''$.

Later, Fan {\it et al.} investigated different TD shapes and their influence on the TE properties, including a gaussian distribution and boxcar function \cite{Fan2011}. They determined that when the integral of the TD is set to a constant the Mahan-Sofo delta function gives the highest $ZT$, and that for a bounded TD a finite-width boxcar function gave the largest $ZT$, which was further evidenced through a genetic algorithm search. Zhou {\it et al.} explored the effect of bandwidth on $ZT$, using a tight-binding model and comparing different scattering models and the role of dimensionality \cite{Zhou2011}, and found that a finite bandwidth provides the optimal $ZT$. Moreover, Zhou {\it et al.} pointed out that the TD remains finite as the bandwidth approaches zero, resulting in a zero $\sigma$ and $ZT$, and thus concluding that the optimal $ZT$ cannot arise from an extremely narrow band. Jeong {\it et al.} further investigated the effect of different bandstructures and scattering models, and concluded that a finite bandwidth produces a higher $ZT$ when $\kappa_l$ is finite \cite{Jeong2012}. More recently, other studies have continued to explore what are the best bandstructures and scattering profiles to enable the optimal TE performance \cite{Xi2016,Kumarasinghe2019,Deng2020,Park2021}.

Given that real materials cannot produce a diverging delta function TD, as pointed out by Mahan and Sofo \cite{Mahan1996}, and that previous studies suggest a finite bandwidth is better, a question that remains is: ``What is the optimal TD for thermoelectrics, when the TD is not permitted to diverge (i.e., {\it bounded})?'' This work focuses on answering this question, and establishing a practical upper limit on the TE figure of merit $ZT$ and $PF$.

\section{Theoretical approach}
\begin{figure}	
	\includegraphics[width=7cm]{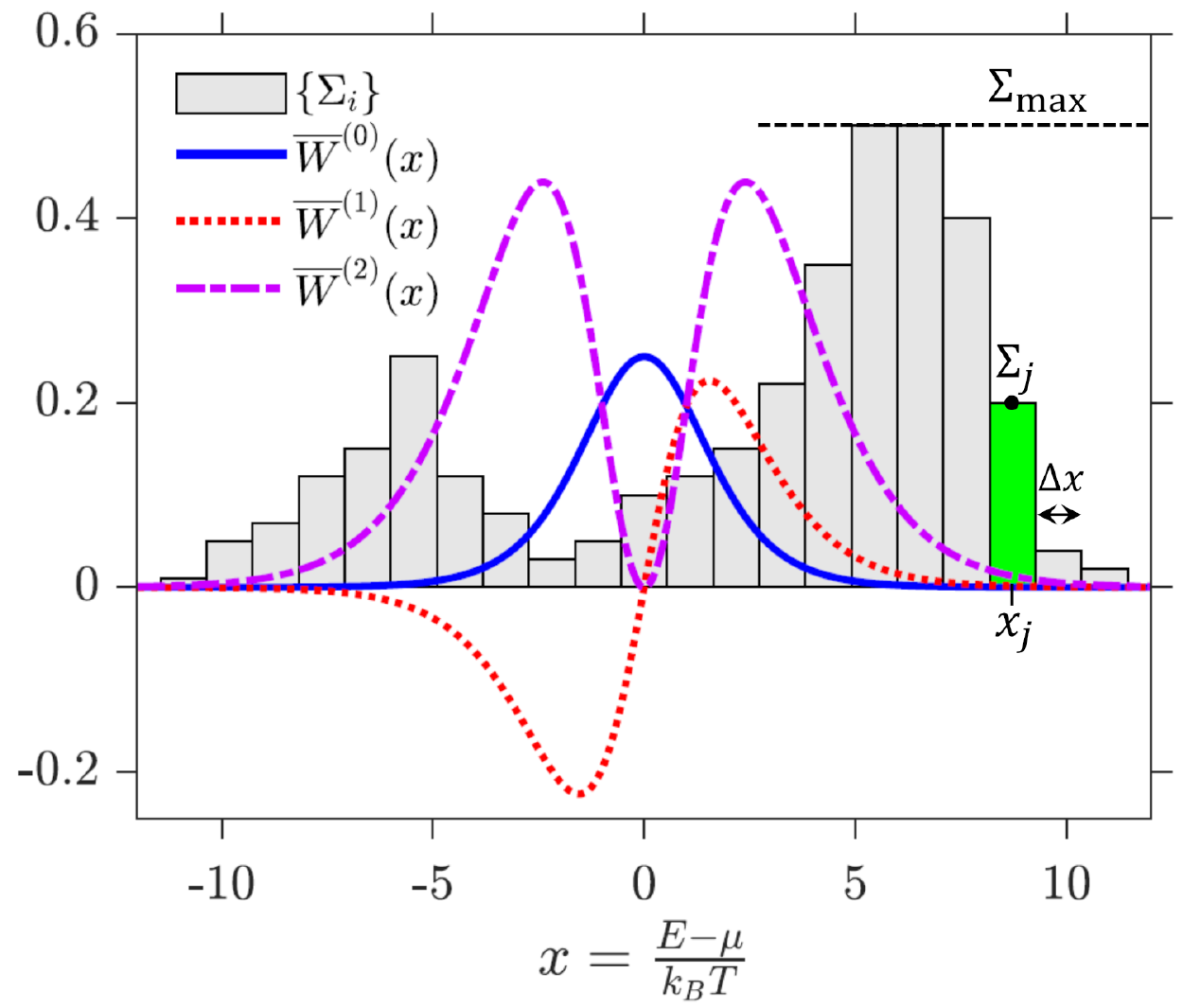} 
	\caption{Discretized transport distribution $\{\Sigma_i\}$ and unitless Fermi window functions $\overline{W}^{(\alpha)}(x)$ versus normalized energy $x$\,=\,$(E-\mu)/k_BT$.}
	\label{fig:binTD_window}
\end{figure}

Assuming a uniform material under near-equilibrium conditions, a solution of the linear Boltzmann transport equation within the relaxation time approximation (RTA) defines the following TE properties:
\begin{align}
\sigma &=  e^2 I_0, \label{eq:cond} \\  
S &= -\left( \frac{k_B}{e} \right) \frac{I_1}{I_0}, \label{eq:seebeck} \\ 
\kappa_e &= k_B^2 T \left( I_2 - \frac{I_1^2}{I_0} \right), \label{eq:kappae}
\end{align}
where $e$\,$>$\,0 is the electron charge magnitude \cite{Thonhauser2003,Jeong2010}. Here, $\kappa_l$ is treated as a constant. The $I_{\alpha}$ are integrals defined as
\begin{align}
I_{\alpha} &= \int_{-\infty}^{\infty} \Sigma(x) \, \overline{W}^{(\alpha)} (x)\, dx, \label{eq:Iint}
\end{align}
where $\Sigma(x)$ is the TD, $\overline{W}^{(\alpha)}(x)$\,=\,$x^{\alpha}[-df_0/dx]$ is the unitless Fermi window function of order $\alpha$ and $f_0$ is the Fermi-Dirac distribution. Fig.~\ref{fig:binTD_window} shows $\overline{W}^{(\alpha)}(x)$ for $\alpha$\,=\,0, 1, 2. $\Sigma(x)$ is the only unknown function in Eqns.~(\ref{eq:cond})-(\ref{eq:Iint}) and represents the central quantity that determines the TE parameters. The TD is expressed as \cite{Mahan1996,Jeong2010}
\begin{align}
\Sigma(E) = \frac{1}{\Omega}\sum_{\bf k} v_z^2({\bf k}) \, \tau({\bf k}) \, \delta(E-\epsilon({\bf k})),
\label{eq:td} 
\end{align}
where $\Omega$ is the sample volume, $\bf k$ are the electron states (includes band index and spin), $\epsilon({\bf k})$ are the electron energies, $v_z({\bf k})$\,=\,$(1/\hbar)[\partial \epsilon/\partial k_z]$ is the velocity along the transport direction (here taken as $z$), and $\tau({\bf k})$ is the scattering time. Eq.~(\ref{eq:td}) can also be written as $\Sigma(E)$\,=\,$(2/h)M(E)\lambda(E)$, where $M(E)$ is the distributions of modes and $\lambda(E)$ is the mean-free-path for backscattering \cite{Rudderham2021}. The TD includes all the detailed electronic properties that are unique to each material.

We begin by expressing the TD as $\Sigma(x)$\,=\,$\psi^2(x)$ and solving for $\psi(x)$, which guarantees that $\Sigma(x)$ is positive. Next, we discretize the normalized energy $x$\,$\rightarrow$\,$\{x_i\}$ and the TD $\Sigma(x)$\,$\rightarrow$\,$\{\Sigma_i\}$\,=\,$\{\psi_i^2\}$, where $i$ runs over all integers and a uniform spacing $\Delta x$ is assumed (see Fig.~\ref{fig:binTD_window}). Note that the $\Delta x$\,$\rightarrow$\,0 limit will be taken later. In this case, the integral Eq.~(\ref{eq:Iint}) can be approximated as
\begin{align}
I_{\alpha} &\approx \Delta x \sum_{i=-\infty}^{\infty} \psi_i^2 \,  \overline{W}_i^{(\alpha)}.  \label{eq:Iint2}
\end{align}
The unknown function $\psi(x)$ is now transformed into $\{\psi_i\}$, an infinite number of unknown variables to be determined. The optimal $\{\psi_i\}$ that maximize $ZT$ or $PF$ can be found from the stationary point conditions $\partial(ZT)/\partial\psi_j$\,=\,0 and $\partial(PF)/\partial\psi_j$\,=\,0, where $\psi_j$\,$\in$\,$\{\psi_i\}$. Before attempting to solve this problem, we introduce a constraint on the solution.

To ensure that the optimized TD is bounded, we impose the condition $R(\{\Sigma_i\}$$-$$\Sigma_{\rm max})=0$, where $R(y)$ is the ramp function (equal to $y$ when $y\ge0$, and equal to zero when $y<0$). It is easy to verify that $R(\{\Sigma_i\}$$-$$\Sigma_{\rm max})=0$ is only satisfied when $\Sigma_i$\,$\le$\,$\Sigma_{\rm max}$, where $\Sigma_{\rm max}$ is some maximum value. This constraint is introduced into the stationary point conditions using Lagrange multipliers, which results in the following functionals to optimize:
\begin{align}
\mathcal{L}_{ZT}[\{\psi_i\}] &= ZT[\{ \psi_i \}] - \sum_i \lambda_i R(\psi_i^2-\psi_{\rm max}^2),  \label{eq:fzt} \\
\mathcal{L}_{PF}[\{\psi_i\}] &= PF[\{ \psi_i \}] - \sum_i \nu_i R(\psi_i^2-\psi_{\rm max}^2),  \label{eq:fpf} 
\end{align}
where $\lambda_i$ and $\nu_i$ are the Lagrange multipliers (one for each $\psi_i$). The solution comes from solving the stationary points of $\mathcal{L}_{ZT}$ or $\mathcal{L}_{PF}$:
\begin{align}
\frac{\partial \mathcal{L}[\{\psi_i\}]}{\partial \psi_j} &= 0.  \label{eq:optztpf} 
\end{align}
One equation is obtained for each of the $\psi_j$ in $\{\psi_i\}$, leading to an infinite number of coupled equations. Fortunately, it is possible to solve for the $\{\psi_i\}$ by focusing on the generic form of only one of these equations. In addition to Eq.~(\ref{eq:optztpf}), solutions must also satisfy $\partial \mathcal{L}_{ZT} / \partial \lambda_j$\,=\,0 and $\partial \mathcal{L}_{PF} / \partial \nu_j$\,=\,0, which simply enforces the condition that $\psi_j$\,$\le$\,$\psi_{\rm max}$.

\section{Results}

\subsection{Optimal transport distribution for power factor}
Eqns.~(\ref{eq:fpf}) and (\ref{eq:optztpf}) are used to derive the optimal bounded TD that maximizes the power factor. Using the property $\partial I_{\alpha} / \partial \psi_j$\,=\,$2\Delta x \,\psi_j \overline{W}_j^{(\alpha)}$ and that the derivative of a ramp function is a Heaviside function ($dR(y)/dy=\Theta(y)$), after some manipulation, Eqns.~(\ref{eq:fpf})-(\ref{eq:optztpf}) become
\begin{align}
\psi_j \bigg\{ &  2 e k_B \Delta x \,S \,\overline{W}_j^{(1)}  + e^2 \Delta x \,S^2\, \overline{W}_j^{(0)} \nonumber \\ & + \nu_j \, \Theta(\psi_j^2 - \psi_{\rm max}^2) \bigg\} = 0.  \label{eq:optpf2}
\end{align}
This equation looks the same for all $\psi_j$. The energy-dependent quantities in Eq.~(\ref{eq:optpf2}) carry a subscript $j$, corresponding to their value at a particular $x_j$. Upon inspection Eq.~(\ref{eq:optpf2}) admits two types of solutions; either $\psi_j=0$ or the factor in curly brackets is zero (or both).

As an intermediate step, to show how the Lagrange multipliers relate to the optimal $PF$, we multiply Eq.~(\ref{eq:optpf2}) by $\psi_j$ and sum over all $j$. Using the definitions for $\sigma$ and $S$, we arrive at
\begin{align}
S^2 \sigma = \sum_j \nu_j \, \psi_j^2 \, \Theta(\psi_j^2 - \psi_{\rm max}^2). \label{eq:pf_lagrange}
\end{align} 
Since $PF$ is always positive, along with all the quantities in Eq.~(\ref{eq:pf_lagrange}) except $\nu_j$, this informs us that $S^2\sigma$ is maximized with the largest positive $\nu_j$ values. This fact will become useful in the next step. We note that the TE parameters appearing in Eqns.~(\ref{eq:optpf2})-(\ref{eq:pf_lagrange}) correspond to the optimal values that maximize $PF$.

\begin{figure}	
	\includegraphics[width=8cm]{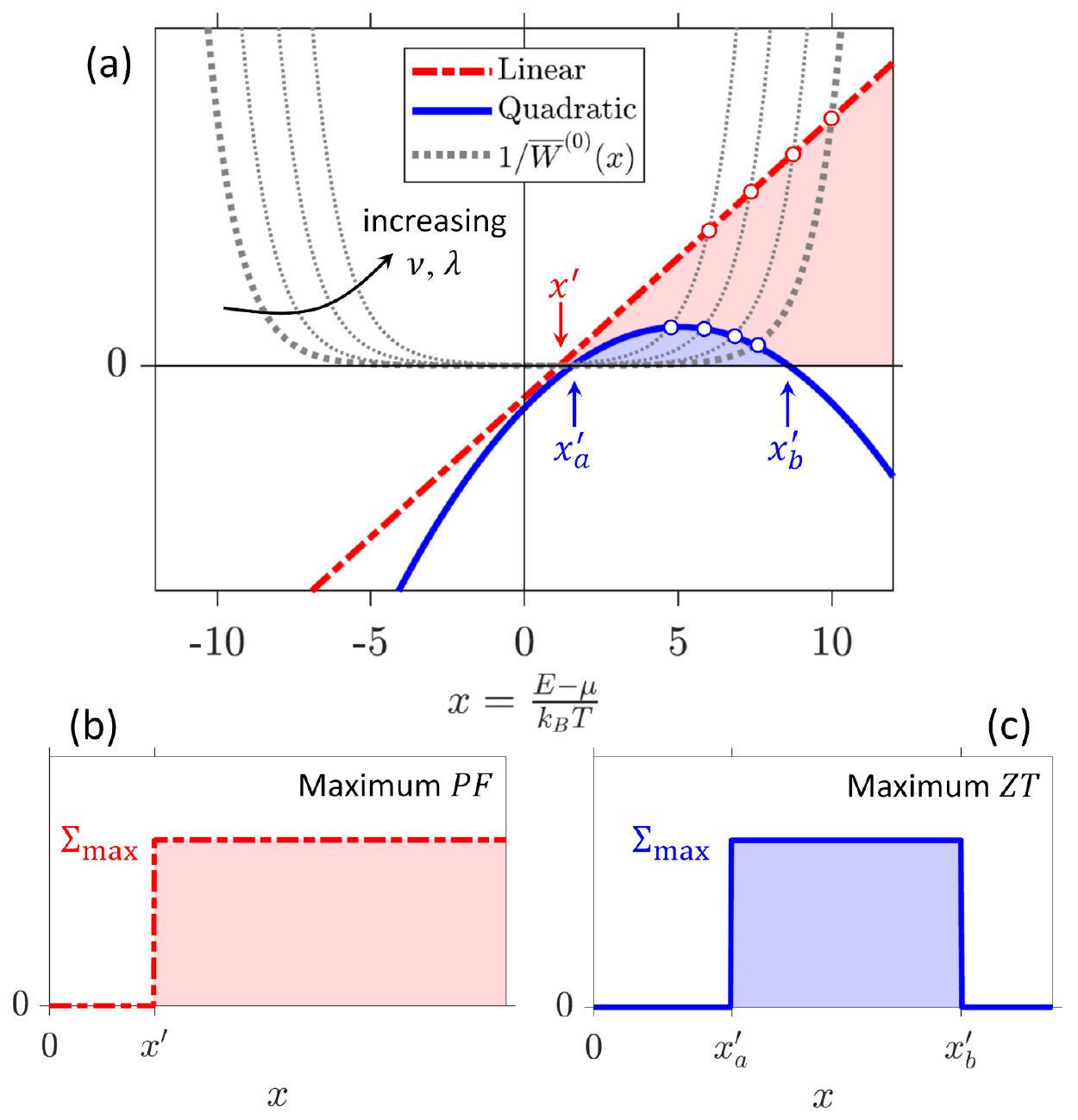} 
	\caption{(a) Linear and quadratic functions of $x$ given by the left-hand side of Eqns.~(\ref{eq:pf_condition}) and (\ref{eq:zt_condition}), respectively, and inverse unitless Fermi window function of zeroth order. Optimal transport distribution for maximum (b) power factor and (c) TE figure of merit, where $x'$\,=\,1.145 and $x'_{a,b}$ are shown in Fig.~\ref{fig:te_props}(a).}
	\label{fig:lin_qua_funcs}
\end{figure}

Next, we investigate the solutions of Eq.~(\ref{eq:optpf2}) for which $\psi_j$\,$\neq$\,0 that arise from setting the factor is curly brackets to zero. First, any solution for which $\psi_j$\,$<$\,$\psi_{\rm max}$ can be eliminated since, according to Eq.~(\ref{eq:pf_lagrange}), this would not contribute to increasing $PF$. Second, we consider the case when $\psi_j$\,=\,$\psi_{\rm max}$. Eq.~(\ref{eq:optpf2}) gives
\begin{align}
-2 e k_B \Delta x \,S\,x_j - e^2 \Delta x \,S^2 = \nu_j / \overline{W}_j^{(0)}. \label{eq:pf_condition}
\end{align}
For simplicity a negative Seebeck coefficient ($S$\,=\,$-|S|$) is assumed, corresponding to an $n$-type TE (the equivalent case of a positive $S$, $p$-type TE, is easily obtained later). The energies $x_j$ that satisfy Eq.~(\ref{eq:pf_condition}) indicate where $\psi_j$\,=\,$\psi_{\rm max}$. The left-hand side is a linear function with positive slope and negative y-intercept, and the right-hand side is a symmetric function that grows exponentially for large $|x_j|$\,$\gg$\,1 (see Fig.~\ref{fig:lin_qua_funcs}(a)). Any $x_j$ can be a solution of Eq.~(\ref{eq:pf_condition}) with the appropriate, and unique, choice in Lagrange multiplier $\nu_j$. However, as noted earlier, Eq.~(\ref{eq:pf_lagrange}) informs us that $PF$ is maximized with the largest positive $\nu_j$. Thus, the maximum $PF$ is obtained by only considering {\it positive} $\nu_j$, which correspond to the $x_j$ for which the left-hand side of Eq.~(\ref{eq:pf_condition}) is greater than zero: $x_j>e|S|/2k_B$.

Letting $\Delta x$\,$\rightarrow$\,0, the optimal TD for power factor has the shape of a Heaviside function:
\begin{align}
\Sigma_{PF}(x) &= \Sigma_{\rm max} \, \Theta(x-x'), \label{eq:td_pf} 
\end{align}
where $x'=e|S|/2k_B$, as shown in Fig.~\ref{fig:lin_qua_funcs}(b). To determine the values of both $x'$ and $S$, $\Sigma_{PF}(x)$ is used to evaluate Eqns.~(\ref{eq:seebeck}) and (\ref{eq:Iint}), resulting in the formula $S$\,=\,$(-k_B/e)[x'+\int_{x'}^{\infty} f_0(x)\,dx/f_0(x')]$. Substituting this expression for $|S|$ into the above condition for $x'$ one obtains $\int_{x'}^{\infty}f_0(x)\,dx$\,=\,$x'\,f_0(x')$, which has the solution $x'$\,=\,1.145 (obtained numerically using binary search). This $\Sigma_{PF}$ results in the following TE parameters: 
\begin{align}
\sigma / (\Sigma_{\rm max}\,k_B^2) &= 3.25\times10^7, &[{\rm K}^2/{\rm V}^2]  \label{eq:pfmax_cond} \\
S &= -197\times 10^{-6}, &[{\rm V}/{\rm K}] \label{eq:pfmax_seebeck} \\ 
PF / (\Sigma_{\rm max}\,k_B^2) &= 1.27, &[{\rm unitless}] \label{eq:pfmax} \\
\kappa_e / (\Sigma_{\rm max}\,k_B^2 T) &= 0.276. &[{\rm unitless}] \label{eq:pfmax_ke}
\end{align}
$\sigma$, $PF$ and $\kappa_e$ all scale linearly with $\Sigma_{\rm max}$, which has units of 
J$^{-1}$m$^{-1}$s$^{-1}$ (for a 3D bulk TE). Note that these results are independent of temperature, and that this particular normalization of the TE parameters is discussed later.

\subsection{Optimal transport distribution for $ZT$}
Next, we find the optimal TD that maximizes $ZT$. Starting with Eqns.~(\ref{eq:fzt}) and (\ref{eq:optztpf}), one finds the following optimization condition:
\begin{align}
\psi_j \bigg\{ & k_B^2 T \Delta x  \frac{ZT}{(\kappa_e+\kappa_l)} \overline{W}_j^{(2)} +  2 e k_B \Delta x  \frac{ZT(1+ZT)}{S \sigma} \overline{W}_j^{(1)}  \nonumber \\
&+ e^2 \Delta x  \frac{ZT(1+ZT)}{\sigma}   \overline{W}_j^{(0)}  + \lambda_j  \Theta(\psi_j^2 - \psi_{\rm max}^2) \bigg\} = 0.  \label{eq:optzt2}
\end{align}
The TE parameters appearing in Eq.~(\ref{eq:optzt2}) correspond to those that maximize $ZT$. Once again, the optimization condition allows for $\psi_j$\,$\neq$\,0 only for the $x_j$ that make the factor in curly brackets vanish.

Before solving Eq.~(\ref{eq:optzt2}), we multiply it by $\psi_j$ and sum over $j$, which allows us to obtain the following equation relating $ZT$ to the Lagrange multipliers $\lambda_j$:
\begin{align}
ZT &= \left( \frac{\kappa_e + \kappa_l}{\kappa_l} \right) \sum_j \lambda_j \, \psi_j^2 \, \Theta(\psi_j^2 - \psi_{\rm max}^2). \label{eq:zt_lagrange}
\end{align}
This expression indicates that $ZT$ is maximized with the largest positive $\lambda_j$ values.

Returning to Eq.~(\ref{eq:optzt2}), we seek the solutions for which $\psi_j$\,$\neq$\,0. First, solutions for which $\psi_j$\,$<$\,$\psi_{\rm max}$ can be omitted as Eq.~(\ref{eq:zt_lagrange}) indicates they would not contribute to $ZT$. Second, when $\psi_j$\,$=$\,$\psi_{\rm max}$ Eq.~(\ref{eq:optzt2}) becomes
\begin{align}
& -k_B^2 T \Delta x  \frac{ZT}{(\kappa_e+\kappa_l)} x_j^2 -  2 e k_B \Delta x  \frac{ZT(1+ZT)}{S \sigma} x_j  \nonumber \\
&- e^2 \Delta x  \frac{ZT(1+ZT)}{\sigma}  =  \frac{\lambda_j}{\overline{W}_j^{(0)}}.  \label{eq:zt_condition}
\end{align}
A negative Seebeck coefficient is assumed $S$\,=\,$-|S|$ for simplicity; the positive $S$ case is straightforwardly obtained later. The $x_j$ that obey Eq.~(\ref{eq:zt_condition}) tell us where $\psi_j$\,$=$\,$\psi_{\rm max}$, with all other $\psi_j$ equal to zero. The right-hand side of Eq.~(\ref{eq:zt_condition}) is the same as the one encountered in Eq.~(\ref{eq:pf_condition}). The left-hand side is a quadratic equation (shown Fig.~\ref{fig:lin_qua_funcs}(a)) with positive vertex, positive local maximum, negative y-intercept and zeros ($x'_{a,b}$):
\begin{align}
x'_{a,b} & = \frac{e}{k_B} \frac{|S|}{ZT} \left[ 1+ZT \pm \sqrt{1+ZT} \right]. \label{eq:zeros}
\end{align}
Given the appropriate choice in Lagrange multiplier $\lambda_j$, all $x_j$ can satisfy Eq.~(\ref{eq:zt_condition}). However, from Eq.~(\ref{eq:zt_lagrange}) we learned that $ZT$ is maximized with the largest positive $\lambda_j$. Hence, we obtain the optimal $ZT$ by considering only $x_j$ solutions that arise for {\it positive} $\lambda_j$. This corresponds to the region where the quadratic equation is positive, delimited by the zeros: $x'_a$\,$\le$\,$x_j$\,$\le$\,$x'_b$.

We conclude that, in the limit $\Delta x$\,$\rightarrow$\,0, the optimal TD for $ZT$ is a boxcar function:
\begin{align}
\Sigma_{ZT}(x) &= \Sigma_{\rm max} \, \big[ \Theta(x-x'_a) - \Theta(x-x'_b) \big], \label{eq:td_zt}
\end{align}
where $x'_{a,b}$ are given by Eq.~(\ref{eq:zeros}), as shown in Fig.~\ref{fig:lin_qua_funcs}(c). Manipulating Eq.~(\ref{eq:zeros}) one can arrive at the following expressions for the optimal $ZT$ and $S$:
\begin{align}
ZT &= \left( \frac{x'_a + x'_b}{x'_a - x'_b} \right)^2 - 1, \,\,\,\, S = \frac{2k_B}{e} \left( \frac{x'_a x'_b}{x'_a + x'_b} \right). \label{eq:zt_s_zeros}
\end{align}
To determine $x'_{a,b}$, we start with an initial guess for $x'_{a,b}$ and calculate $ZT$ and $S$ using Eqns.~(\ref{eq:cond})-(\ref{eq:Iint}) assuming a boxcar TD, then insert these values into Eq.~(\ref{eq:zeros}) to obtain updated $x'_{a,b}$. This process is repeated until the absolute relative differences in $ZT$ and $S$ calculated using Eqns.~(\ref{eq:cond})-(\ref{eq:Iint}) versus Eq.~(\ref{eq:zt_s_zeros}) reach below 10$^{-10}$.

\begin{figure*}	
	\includegraphics[width=17cm]{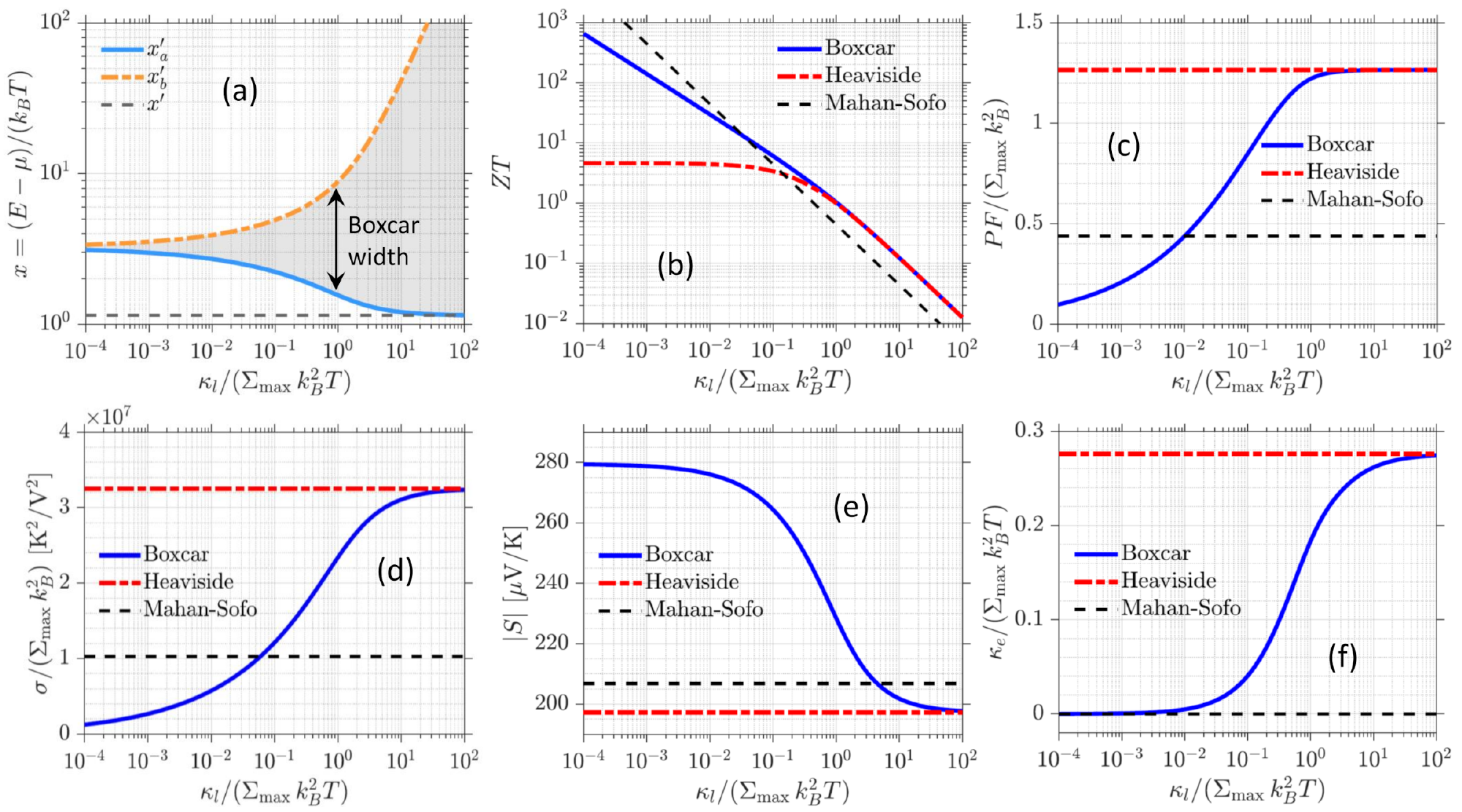}
	\caption{(a) Edges of the boxcar TD, $x'_{a,b}$, and edge of the Heaviside TD, $x'$. (b) TE figure of merit $ZT$, (c) normalized power factor $PF$, (d) normalized electrical conductivity $\sigma$, (e) Seebeck coefficient $|S|$, (f) normalized electronic thermal conductivity $\kappa_e$ versus normalized lattice thermal conductivity $\kappa_l$. Panels (b)-(f) show the results of the optimal boxcar and Heaviside TDs, along with the Mahan-Sofo limit (plotted versus $\kappa_l/(\Sigma_{\rm MS} \, k_B^2T)$).} \label{fig:te_props}
\end{figure*}

\subsection{Thermoelectric properties of the optimal transport distributions}
When calculating $ZT$ it is convenient to redefine the TE parameters. In order to create unitless quantities, starting from the definition of $ZT$, we divide by $\Sigma_{\rm max} k_B^2 T$ in both the numerator and denominator to arrive at:
\begin{align}
ZT &= \frac{S^2 \overline{\sigma}}{\big( \overline{\kappa}_e + \overline{\kappa}_l \big) }, \label{eq:zt_alter}
\end{align}
where $\overline{\sigma}$\,=\,$\sigma$/($\Sigma_{\rm max} k_B^2)$, $\overline{\kappa}_e$\,=\,$\kappa_e$/$(\Sigma_{\rm max} k_B^2 T)$, and $\overline{\kappa}_l$\,=\,$\kappa_l$/$(\Sigma_{\rm max} k_B^2 T)$. In this form $\overline{\sigma}$, $S$, $\overline{\kappa}_e$ are independent of $\Sigma_{\rm max}$ and $T$, and $\overline{PF}$\,=\,$S^2\overline{\sigma}$, $\overline{\kappa}_e$ and $\overline{\kappa}_l$ are unitless. Moreover, the effects of $\Sigma_{\rm max}$ and $T$ on $ZT$ are conveniently captured by the $\overline{\kappa}_l$ term.

The calculated $x'_{a,b}$ versus $\overline{\kappa}_l$\,=\,$\kappa_l/(\Sigma_{\rm max} k_B^2T)$ are presented in Fig.~\ref{fig:te_props}(a). The boxcar width, $x'_b-x'_a$, increases with $\overline{\kappa}_l$. For small $\overline{\kappa}_l$, both $x'_{a,b}$ approach a value of roughly 3.24. For large $\overline{\kappa}_l$, $x'_a$ tends towards $x'$\,=\,1.145 as $x'_b$ increases rapidly, thus retrieving the optimal Heaviside TD. Figures~\ref{fig:te_props}(b)-(c) show the figure of merit and power factor, for both the optimal boxcar and Heaviside TDs. As expected, the boxcar TD provides an upper limit on $ZT$, and the Heaviside TD provides an upper limit on $PF$. The boxcar $ZT$ increases without limit when reducing $\overline{\kappa}_l$, with $ZT$\,$\ge$\,1 requiring $\overline{\kappa}_l$\,$\le$1. An important consequence of achieving a $ZT$ greater than roughly 10 is the strong reduction in $PF$ compared to its upper limit.

When $\overline{\kappa}_l$\,$\gg$\,1, both the boxcar and the Heaviside TDs give near identical $ZT$. In this regime, where $\overline{\kappa}_l$\,$\gg$\,$\overline{\kappa}_e$ (see Fig.~\ref{fig:te_props}(f) and the Supplemental Material \cite{Suppl}), $ZT$ is maximized by adopting the largest possible power factor. Fig.~\ref{fig:te_props}(c) shows that $PF$ approches its maximum value, which explains why the boxcar TD asymptotically approaches the Heaviside TD as $\overline{\kappa}_l$\,$\rightarrow$\,$\infty$.

When $\overline{\kappa}_l$\,$\ll$\,1, the $ZT$ originating from the boxcar TD splits from that of the Heaviside TD. With the latter, the constant $\overline{\kappa}_e$ dominates over $\overline{\kappa}_l$, resulting in a $ZT$ that saturates to a value of $\approx$\,4.6. With the former, $\overline{\kappa}_e$  decreases to allow $ZT$ to grow, at the expense of a lower power factor. One finds that $\overline{\kappa}_e$\,$\rightarrow$\,$\overline{\kappa}_l$\,/2 as $\overline{\kappa}_l$ decreases (see the Supplemental Material \cite{Suppl}). Since $\kappa_e$ is directly related to the variance of the transport distribution \cite{Mahan1996}, decreasing $\overline{\kappa}_e$ requires decreasing the boxcar width, as seen in Fig.~\ref{fig:te_props}(a). This also lowers $\overline{\sigma}$ and $\overline{PF}$. The Seebeck coefficient, however, increases as the boxcar narrows because $x'_a$ rises, thus increasing the average energy of current flow. While $\overline{\sigma}$, $\overline{\kappa}_e$ and $\overline{PF}$ vanish as $\overline{\kappa}_l$\,$\rightarrow$\,0, and $ZT$ might be expected to go to zero \cite{Zhou2011}, $ZT$ increases without limit since the Lorenz number $\kappa_e/(\sigma T)$ tends to zero (see the Supplemental Material \cite{Suppl}). 

The results shown have been for the case of a negative $S$ ($n$-type TE); the positive $S$ case ($p$-type TE) is obtained by applying a reflection operation about $x$\,=\,0 on the optimal TDs, $\Sigma(x)$\,$\rightarrow$\,$\Sigma(-x)$.

\section{Discussion}
For comparison, the results of the Mahan-Sofo limit \cite{Mahan1996} are shown in Fig.~\ref{fig:te_props}(b)-(f). While the boxcar TD yields higher $ZT$ for large $\overline{\kappa}_l$ due to its $PF$, and the Mahan-Sofo limit gives higher $ZT$ for small $\overline{\kappa}_l$ due to its zero $\kappa_e$, such a comparison assumes $\Sigma_{\rm max}$\,=\,$\Sigma_{\rm MS}$ which may not be meaningful. If the prefactors are chosen such that $\sigma$ is made to agree, the $ZT$s are nearly the same (see the Supplemental Material \cite{Suppl}).

As discussed above, the key parameter controlling the maximum $ZT$ is $\overline{\kappa}_l$\,=\,$\kappa_l$/$(\Sigma_{\rm max} k_B^2 T)$. For a given $\overline{\kappa}_l$ the upper limit $ZT$ corresponds to the largest $ZT$, or TE efficiency, that is possible with a bounded transport distribution. To increase $ZT$ one must lower $\overline{\kappa}_l$, which is achieved by decreasing $\kappa_l$, increasing $\Sigma_{\rm max}$, or increasing $T$. Since $\sigma$\,$\propto$\,$\Sigma_{\rm max}$, this is effectively a statement of the ``phonon-glass, electron-crystal'' concept \cite{Slack1995}. There are several successful strategies to reduce $\kappa_l$ that have been demonstrated, including alloying \cite{Khatami2016}, nanostructuring \cite{Poudel2008,Biswas2012}, high anharmonicity \cite{Zhao2014}, among others. Large $\Sigma_{\rm max}$ is possible in materials that possess a large distribution of modes, high velocities and low electron scattering. We note that $1/\overline{\kappa}_l$ is closely related to the generalized $b$-factor \cite{Witkoske2019}. For a typical temperature dependency of $\kappa_l$\,$\propto$\,$1/T$ (phonon-phonon scattering \cite{Ward2010}) and $\Sigma_{\rm max}$\,$\propto$\,$1/T$ (electron-phonon scattering \cite{Lundstrom2000}), one would expect $\overline{\kappa}_l$ to scale as $1/T$.

What is a typical value of $\overline{\kappa}_l$ for an established good TE? From its definition, $\overline{\kappa}_l$\,=\,$\kappa_l$/$(\Sigma_{\rm max} k_B^2 T)$, it is clear that $\overline{\kappa}_l$ depends on both phonon and electron transport properties, since it is proportional to $\kappa_l$ and inversely proportional to $\Sigma_{\rm max}$. Most bulk TEs tend to have relatively smooth transport distributions that increase with energy away from the band edge, however $\Sigma_{\rm max}$ can be estimated by using the TD value at roughly 10$k_B T$ from the band edge (beyond which the TD is suppressed exponentially by the $\overline{W}^{(\alpha)}$). For SnSe, a TE with $ZT$\,$\approx$\,2.8 \cite{Chang2018}, we estimate $\Sigma_{\rm max}$\,$\sim$\,2$\times$10$^{44}$\,J$^{-1}$m$^{-1}$s$^{-1}$ from Refs.~\cite{Kutorasinski2015,Ma2018}. With a room temperature $\kappa_l$\,$\approx$\,0.7\,Wm$^{-1}$K$^{-1}$ \cite{Zhao2014}, we obtain a value of $\overline{\kappa}_l$\,$\sim$\,0.06, which has an upper limit $ZT$ of roughly 9 with an associated $PF$ of $\approx$280~$\mu$Wcm$^{-1}$K$^{-2}$. For Bi$_2$Te$_3$, a TE with $ZT$ below 1 \cite{Witting2019}, we estimate $\Sigma_{\rm max}$\,$\sim$\,5$\times$10$^{43}$\,J$^{-1}$m$^{-1}$s$^{-1}$ \cite{Pettes2013}, which coupled with a room temperature $\kappa_l$\,$\approx$\,1.5\,Wm$^{-1}$K$^{-1}$ \cite{Goldsmid1958} gives $\overline{\kappa}_l$\,$\sim$\,0.5. In this case, the upper limit $ZT$ is approximately 2 with a $PF$ of $\approx$110~$\mu$Wcm$^{-1}$K$^{-2}$. These upper limit values indicate what could be possible with SnSe or Bi$_2$Te$_3$ if their transport distributions had the optimal shape, without any change to their $\Sigma_{\rm max}$.

Why is a boxcar or Heaviside shape optimal for $ZT$ and $PF$? $\Sigma(x)$ is related to a material's capacity to conduct electrons at an energy $x$\,=\,$(E-\mu)/k_B T$, while $\overline{W}^{(0)}(x)$ accounts for how the states are occupied (specifically, the nonequilibrium component of the electron distribution). As a result, $\sigma$ grows when the product of $\Sigma(x)$ and $\overline{W}^{(0)}(x)$ is large. $|S|$, however, increases when $\Sigma(x)$ overlaps more with either the positive or negative side of $\overline{W}^{(1)}(x)$ (see Fig.~\ref{fig:binTD_window}). This is because the Seebeck coefficient is proportional to the average energy of electron flow relative to the chemical potential, which benefits from having conduction occur purely (and far) above or below $\mu$. The discontinuous edge of the optimal Heaviside TD allows both $\sigma$ and $|S|$ to be enhanced, thus maximizing the power factor, compared to a more slowly varying function. For $ZT$ to be maximized one must also seek to decrease $\kappa_e$, without drastically reducing the power factor. As discussed by Mahan and Sofo \cite{Mahan1996} the electronic thermal conductivity is proportional to the variance of $\Sigma(x)\,\overline{W}^{(0)}(x)$; using Eqns.~(\ref{eq:cond})-(\ref{eq:Iint}) one can show that $\kappa_e/\sigma$\,=\,$(k_B^2T/e^2)[\langle x^2 \rangle$$-$$\langle x \rangle^2]$, as well as $S$\,=\,$-$$(k_B/e)\langle x \rangle$. The finite width of the boxcar TD has a lower variance, and thus lower $\kappa_e$, compared to the Heaviside TD, but retains the sharp edges that lead to high power factor. When these sharp features are smoothed out, $ZT$ and $PF$ can be significantly degraded (see the Supplemental Material \cite{Suppl}).

While the optimal TDs derived in this study are very different from the delta function TD obtained by Mahan and Sofo, they are derived following the same underlying principles. The best TD is one that maximizes the absolute average of the distribution $\Sigma(x)\,\overline{W}^{(0)}(x)$ (largest $|S|$) and minimizes its variance (lowest $\kappa_e$) for a given conductivity. For a TD with unrestricted magnitude the best case corresponds to a delta function, for which $\kappa_e$\,$\rightarrow$\,0 while $\sigma$ and $S$ remain finite. For a bounded TD the best case is a boxcar function with a finite width, in order for $\sigma$ to be non-zero, which results in a non-zero $\kappa_e$ that is as small as needed to maximize $ZT$. The optimization principles in this work and that of Mahan and Sofo are fundamentally the same, with the difference in the solutions originating from the constraint imposed on the TD.

The findings from this work indicate there are two main strategies for approaching the TE limits: 1) Increase the magnitude of the TD, $\Sigma_{\rm max}$, which can be achieved by increasing the distribution of modes and/or the mean-free-path for backscattering. This point is somewhat obvious, as it states that a more conductive TE is better. 2) Find or design TE materials with the optimal boxcar or Heaviside shape, the edges of which may need to be tuned according to the particular value of $\kappa_l/(\Sigma_{\rm max} k_B^2 T)$. The optimal TDs can serve as targets when exploring the potential of novel TE materials or band engineering strategies. Establishing what are the optimal bounded TDs, and their associated TE limits, addresses one part of the broader challenge of developing the best TEs. The other (bigger) part to this challenge is determining what specific band structures and scattering profiles achieve the TE upper limits, and which materials display such properties. The derived optimal TDs are applicable to uniform TEs under near-equilibrium conditions that can be described by the RTA. This includes TEs of different dimensionality (3D, 2D or 1D), as well as transport regimes from diffusive to ballistic. 1D semiconductor nanowires can display a Heaviside or boxcar TD, when assuming a constant mean-free-path or ballistic transport \cite{Hicks1993,Kim2009,Zhou2011,Jeong2012}. In particular, a boxcar TD could be achieved in systems described by a 1D nearest-neighbor tight-binding model \cite{Zhou2011,Jeong2012,Whitney2014} including, for example, a line of quantum dots or an atomic chain wherein the isolated energy levels are broadened due to inter-site coupling. 2D ring-shaped band semiconductors have also shown discontinuous TDs similar to a Heaviside function \cite{Zahid2010,Maassen2013,Wickramaratne2015,Rudderham2021}. Obtaining a 3D bulk TE with the ideal TD shape would be beneficial to avoid the issue of packing fraction \cite{Kim2009}.

The results of this study provide a theoretical explanation to earlier numerical work, based on a genetic algorithm search, that found a boxcar function to be the optimal bounded TD for $ZT$ \cite{Fan2011}. Moreover, this work establishes the precise location of the rising and falling edges of the boxcar function under all conditions, which are controlled by the factor $\kappa_l/(\Sigma_{\rm max} k_B^2 T)$. Lastly, we note the similarity of the optimal TDs in this work to the optimal transmission function derived for quantum TEs \cite{Whitney2014}, which is also a boxcar function.

\section{Conclusion}
The optimal bounded (by some finite value $\Sigma_{\rm max}$) TDs that maximize the TE figure of merit and power factor were derived and determined to be a boxcar and Heaviside function, respectively. These optimal TDs provide theoretical upper limits on $ZT$ and $PF$. While $PF$ has a maximum value determined by $\Sigma_{\rm max}$, the optimal $ZT$ increases without limit as the key quantity $\kappa_l/(\Sigma_{\rm max} k_B^2 T)$ decreases. To achieve the maximum $ZT$ the edges of the boxcar TD must be located at specific energies. These results suggest two main approaches to enhance TE performance, aside from lowering $\kappa_l$, which involve identifying or designing materials with TDs that have large magnitude $\Sigma_{\rm max}$ (requiring a large distribution of modes, high-velocity states and low scattering) and that possess the ideal boxcar or Heaviside shape (controlled by the dispersion shape, scattering profile and dimensionality).

{\it Acknowledgments.}---This work was supported by NSERC (Discovery Grant No. RGPIN-2016-04881) and Compute Canada.


\end{document}